\documentclass[showpacs,superscriptaddress,aps ]{revtex4}
\usepackage{amssymb}
\usepackage[centertags]{amsmath}
\usepackage{txfonts}
\usepackage{epsfig}
\usepackage{bm}
\usepackage{color}
\usepackage{graphicx,graphics}
\usepackage{multirow}
\usepackage{ulem}

\begin{document}

\title{Quark charge balance function and hadronization effects in relativistic heavy ion collisions}

\author{Jun Song}
\affiliation{Department of Physics, Jining University, Jining, Shandong 273155, China}
\affiliation{School of Physics, Shandong University, Jinan, Shandong 250100,  China}
\author{Feng-lan Shao}
\affiliation{Department of Physics, Qufu Normal University, Shandong 273165, China}
\author{Zuo-tang Liang}
 \affiliation{School of Physics, Shandong University, Jinan, Shandong 250100, China}

\begin{abstract}
We calculate the charge balance function of the bulk quark system before hadronization
and those for the directly produced and the final hadron system in high energy heavy ion collisions.
We use the covariance coefficient to describe the strength of the correlation between the momentum of the quark and 
that of the anti-quark if they are produced in a pair and fix the parameter by comparing the results for hadrons with the available data.
We study the hadronization effects and decay contributions by comparing the results for
hadrons with those for the bulk quark system.
Our results show that while hadronization via quark combination mechanism slightly increases the width of the charge balance functions,
it preserves the main features of these functions such as the longitudinal boost invariance and scaling properties in rapidity space.
The influence from resonance decays on the width of the balance function is more significant but it does not
destroy its boost invariance and scaling properties in rapidity space either.
The balance functions in azimuthal direction are also presented.
\end{abstract}

\pacs{25.75.Dw, 25.75.Gz, 25.75.Nq, 25.75.-q}
\maketitle

\section{introduction}

The electric charge balance function for the final state hadrons has been proposed
as a probe to study the properties of the bulk matter system produced in relativistic heavy ion collisions
\cite{Bass_clockHd,SJeon02,Bialas04,ChengS04,Bozek05StatResonance,LNbf09,Pratt11_bf}.
Measurements have already carried out both in the rapidity space \cite{StarBF130,NA49SCden,NA49REden} and in the azimuthal direction\cite{StarBF200AApp}.
From the data now available\cite{StarBF130,StarBF200AApp}, we are already able to see clearly that the charge balance functions for hadrons produced in high energy heavy ion collisions are
significantly narrower than those for  $pp$ collisions at the same energies and they are narrower for central collisions than those for peripheral collisions,
indicating a strong local charge compensation in the bulk quark matter system produced in heavy ion collisions.
The data \cite{StarBF200Scaling} further show that the charge balance functions have the longitudinal boost invariance
and scaling properties in the rapidity space,
and these properties hold for either transverse momentum $p_T$-integrated balance functions or those for different $p_T$ ranges.

These features of the experimental data\cite{StarBF130,StarBF200AApp,StarBF200Scaling} are rather striking and suggest that such studies should be able to give more insights to the understanding
of the properties of the bulk quark matter system produced in $AA$ collisions.
It is thus natural to ask whether such behavior hold also for the quark anti-quark system before hadronization.
It is also important to see how large the influence from the hadronization and resonance decay.

In this paper, we propose a simple working model to calculate the charge balance function for the bulk quark anti-quark system before hadronization.
We introduce the variance coefficient $\rho$ to describe the local  correlation in the momentum distribution for the quark and that for the anti-quark if they are produced in a pair.
The parameter $\rho$ measures the strength of the quark-anti-quark momentum correlation produced in the processes.
We study the influence due to hadronization process including the
contributions due to resonance decay by simulating the hadronization process using a quark combination model which describe the final hadron distributions.

The paper is organized as follows. In Sec. II, we study the charge balance of the  quark system before hadronization.
In Sec. III, we study the charge balance function of initial hadron system as well as final hadron system,
and compare them with that of quark system. Sec. IV gives a brief summary.

\section{Charge balance function of the system of quarks and anti-quarks}

We recall that , the balance function is in general defined as \cite{Bass_clockHd},
\begin{equation}
B (\Delta_{2}|\Delta_{1} )=
\frac{1}{2} \Big\{ \rho (b,\Delta_{2}|a,\Delta_{1} )-\rho (a,\Delta_{2}|a,\Delta_{1} )+\rho (a,\Delta_{2}|b,\Delta_{1} )-\rho (b,\Delta_{2}|b,\Delta_{1} ) \Big\} ,
\label{eq:bfdef}
\end{equation}
where $\rho (b,\Delta_{2}|a,\Delta_{1} )$ is the conditional probability of observing a particle of type $b$ in bin $\Delta_{2}$
given the existence of a particle of type $a$ in bin $\Delta_{1}$.
The label $a$ may e.g. refer to all positively charged particles while $b$ refers to all negatively charged ones;
$a$ may also refer to all particles with strangeness $-1$ while $b$ refers to those with $+1$, and so on.
For a system consisting of many particles, the conditional probability $\rho (b,\Delta_{2}|a,\Delta_{1} )$
is calculated by counting the number  $N (b,\Delta_{2}|a,\Delta_{1} )$ of the $ab$-pairs where
$a$ is in bin $\Delta_1$ and $b$ is in bin $\Delta_2$
and the number $N (a,\Delta_{1} )$ of $a$ in bin $\Delta_{1}$, i.e.,
\begin{equation}
\rho (b,\Delta_{2}|a,\Delta_{1} )=\frac{N (b,\Delta_{2}|a,\Delta_{1} )}{N (a,\Delta_{1} )}.
\end{equation}
These numbers can be calculated using the usual two-particle joint momentum distribution function $f_{ab}(\bm{p}_{1},\bm{p}_{2})$
and single particle distribution function $f_{a}(\bm{p})$ or $f_{b}(\bm{p})$ respectively.
They are given by,
 \begin{equation}
 N(b,\Delta_{2}|a,\Delta_{1} )=\int_{\Delta_{1}}d^{3}p_{1}\int_{\Delta_{2}}d^{3}p_{2} f_{ab}(\bm{p}_{1},\bm{p}_{2}),
 \end{equation}
\begin{equation}
N (a,\Delta_{1})=\int_{\Delta_{1}} d^{3}p_{1} f_{a}(\bm{p}_{1}).
\end{equation}

We see that, if $a$ is locally compensated by $b$, the balance function $B (\Delta_{2}|\Delta_{1} )$ should have a very narrower distribution.
In the opposite case, it should be flat.
In the case that $a$ is globally compensated by $b$, e.g.,  for electric charge balance function where
$a$ and $b$ denote positively or negatively charged particle respectively,
the balance function is normalized to unit, i.e. $\sum_{\Delta_{2}}B (\Delta_{2}|\Delta_{1} )=1$.

\subsection{A working model for the two particle joint momentum distribution  functions in the bulk quark matter system}

We consider the bulk quark matter system produced in heavy ion collisions at high energies.
We suppose that the system is composed of $N_q$ quarks and $N_{\bar q}$ anti-quarks.
We denote the normalized momentum distribution of the quarks and anti-quarks by
$n_q(\bm{p})$ and $n_{\bar q}(\bm{p})$ respectively.
In heavy ion collisions, the bulk matter system consists of new created quarks, anti-quarks
and the quarks from the incident nuclei.
Those quarks from the incident nuclei are referred as the net quarks and the new born quarks and anti-quarks are created in pairs.

To obtain a charge balance function that is narrower than that for the completely uncorrelated case,
we introduce a minimum correlation in the two particle joint momentum distributions in the system.
To this end, we construct the following working model for the two particle joint momentum distribution for the bulk quark matter system.
We assume that there is no correlation between the momentum distributions of two different quarks or two anti-quarks.
The joint distributions are simply the products of the corresponding single particle momentum distributions, i.e.
\begin{eqnarray}
&f_{q_1q_2}(\bm{p}_1,\bm{p}_2)=N_{q_1}N_{q_2}n_{q_1}(\bm{p}_1)n_{q_2}(\bm{p}_2)(1-\delta_{q_1,q_2})+ N_{q_1}(N_{q_2}-1)n_{q_1}(\bm{p}_1)n_{q_2}(\bm{p}_2)\delta_{q_1,q_2}, \\
&f_{\bar q_1\bar q_2}(\bm{p}_1,\bm{p}_2)=N_{\bar q_1}N_{\bar q_2}n_{\bar q_1}(\bm{p}_1)n_{\bar q_2}(\bm{p}_2)(1-\delta_{q_1,q_2})+ N_{\bar q_1}(N_{\bar q_2}-1)n_{\bar q_1}(\bm{p}_1)n_{\bar q_2}(\bm{p}_2)\delta_{q_1,q_2},
\end{eqnarray}
where $q_1$ and $q_2$ denote the flavors of the quarks.
For the $q\bar q$ joint momentum distribution, we introduce a correlation between the moment distribution
of the quark and that of the anti-quark which are produced in the same pair.
In this case, the joint distribution for a quark $q_1$ and an anti-quark $q_2$ is given by,
\begin{equation}
f_{q_1\bar q_2}(\bm{p}_1,\bm{p}_2)=
N_{q_1}N_{\bar{q_2}}n_{{q_1}}(\bm{p}_{1})n_{\bar{q_2}}(\bm{p}_{2}) +
N_{\bar{q_1}} \bigl[ n^{pair}_{q\bar{q}}( \bm{p}_{1}, \bm{p}_{2} )-n_{{\bar q_1}}(\bm{p}_{1})n_{\bar{q_2}}(\bm{p}_{2})\bigr] \delta_{q_1,q_2}.
\label{fqqbar}
\end{equation}
The single particle momentum distributions are related to $n_{q\bar q}^{pair}(\bm{p}_1,\bm{p}_2)$ by,
\begin{eqnarray}
& n_{\bar q}(\bm{p}_2)=\int d^3p_1 n^{pair}_{q\bar q}(\bm{p}_1,\bm{p}_2),\\
& n_q(\bm{p})=\frac{N_{\bar q}}{N_q}n_{\bar q}(\bm{p})+\frac{N_{net}}{N_q}n_{net}(\bm{p}).
\end{eqnarray}
Hence, as long as we know $n^{pair}_{q\bar q}(\bm{p}_1,\bm{p}_2)$ and $n_{net}(\bm{p})$,
we can calculate the two particle joint momentum distributions for $qq$, $\bar q\bar q$ and $q\bar q$-system.

To calculate $n^{pair}_{q\bar q}(\bm{p}_1,\bm{p}_2)$, we adopt the picture of the hydrodynamic theory.
Here, we assume the local thermalization and collectivity in the system \cite{Kolb0305084nuth,StarRev,PhenixRev}.
Hence, in the co-moving frame of the fluid cell, due to local thermalization, we  take a Boltzmann distribution for the single quark or anti-quark distribution, i.e.,
\begin{equation}
n^*_q (\bm{p}^* )=n_{th} (\bm{p}^* )=\frac{1}{4\pi m^{2}T K_{2}(m/T)}e^{-E^*/T },
\end{equation}
where the supscript $*$ denote that these quantities are in the co-moving frame, $K_2$ is the Bessel function,
$m$ is the mass of the constituent quark (340 MeV for $u$ or $d$ quark and 500 MeV for strange quark),
and $E^*=\sqrt{\bm{p}^{*2}+m^2}$ is the energy of quark;
$T$ is the temperature of the system at hadronization (take as $T=165$ MeV  \cite{Karsch2002NPA}).

For the joint momentum distribution of the quark and the anti-quark produced in the same pair,
we use the covariance coefficient $\rho$ to describe the correlation between them.
We recall that for a joint momentum distribution fof a $q\bar q$-system, the covariance coefficient $\rho$ is defined as
$\rho={\rm cov} (\bm{p}_q,\bm{p}_{\bar{q}} )/ {\rm var} (\bm{p}_{\bar{q}} )$,
where  ${\rm cov} (\bm{p}_{q},\bm{p}_{\bar{q}} )\equiv \langle {\bm{p}}_q \cdot{\bm{p}}_{\bar{q}} \rangle-\langle{\bm{p}}_q\rangle\cdot \langle{\bm{p}}_{\bar q}\rangle$
and ${\rm var} (\bm{p}_{\bar q})\equiv \langle {\bm{p}}_{\bar{q}}^2 \rangle-\langle{\bm{p}}_{\bar q}\rangle^2$.
We take the joint distribution for the $q\bar q$-system in the co-moving frame of the $q\bar q$-pair in the
Cholesky  factorization form, i.e.,
\begin{equation}
n^{pair*}_{q\bar{q}} (\bm{p}^*_q,\,\bm{p}^*_{\bar{q}} )=
\frac{1}{2({1-\rho^{2}})^{3/2}} [n_{th} (\bm{p}^*_{q} )n_{th} (\frac{\bm{p}^*_{\bar{q}}-\rho\bm{p}^*_{q}}{\sqrt{1-\rho^{2}}} )+
n_{th} (\bm{p}^*_{\bar{q}} )n_{th} (\frac{\bm{p}^*_{q}-\rho\bm{p}^*_{\bar{q}}}{\sqrt{1-\rho^{2}}} ) ].
\label{eq:covrho}
\end{equation}
The covariance parameter $\rho$ describe the strength of the correlation.
If $\rho=0$, there is no correlation between the momentum distribution of the quark and that of the anti-quark and we obtain the factorized form.
For $\rho$ very close to unity, we get a maximum correlation between the momentum of ${\bm p}_q$ and ${\bm p}_{\bar q}$, where the probability is
non-zero only when ${\bm p}_q={\bm p}_{\bar q}$.
In general $-1\le \rho \le 1$, and $\rho>0$ means short range compensation of $q$ and $\bar q$ while $\rho<0$ means the opposite.

The joint distribution $n^{pair}_{q\bar q}( \bm{p}_{q}, \bm{p}_{\bar{q}})$  in the laboratory frame is obtained from $n^{pair*}_{q\bar q}(\bm{p}^*_q,\,\bm{p}^*_{\bar{q}} )$.
Here, we first make the Lorentz transformation ($\bm{\beta}$) from the co-moving frame of the fluid cell to the laboratory frame
to obtain $ n^{pair}_{q\bar{q}}(\bm{p}_{q},\, \bm{p}_{\bar{q}},\bm{\beta})$,
then sum up the contributions from different fluid cells in the system with different collective velocities, i.e.,
\begin{equation}
n^{pair}_{q\bar{q}}( \bm{p}_{q}, \bm{p}_{\bar{q}})
=\int h(\bm{\beta})\, n^{pair}_{q\bar{q}}(\bm{p}_{q},\, \bm{p}_{\bar{q}},\bm{\beta})\, d^{3}\beta,
\end{equation}
where $h(\bm{\beta})$ is the so-called velocity function which corresponds to the velocity distribution of the fluid cell in the system.

The velocity function $h(\bm{\beta})$ is normalized to unity and can be decomposed into the longitudinal part $h_{L}$ and the transverse part $h_{\perp}$.
The longitudinal velocity $\beta_{z}$ is usually replaced by the rapidity $y$.
The azimuthal dependence is isotropic, we integrate it out and obtain,
$ \int h(\bm{\beta}) \,d^{3} \beta= \int h_{L}(y) h_{\perp}(\beta_{\perp})\,dy d\beta_\perp$.
This velocity function $h(\bm{\beta})$ determines, together with the momentum distribution of the quark and anti-quark in the fluid cell,
the single quark spectrum thus the inclusive momentum distribution of the hadrons after hadronization.
In practice, it is parameterized by fitting the data for the hadron momentum distributions with the aid of hadronization models.
According to the transparency observed in experiments\cite{bearden04stop},  and because of that
the observed rapidity spectra of hadrons show a roughly Gaussian shape in the full rapidity range\cite{BeardMeson04},
we take the transverse part as a uniform distribution between $[0,\beta_{\perp}^{max}]$ and parameterize the longitudinal part in a Gaussian-like form,
\begin{equation}
  h_{L}(y)=\frac{1}{ 2 \sigma^{\frac{2}{a}} \Gamma(1+\frac{1}{a})}  e^{-|y|^{a}/\sigma^2}.
\end{equation}
%
The free parameters $a$, $\sigma$ and $\beta_{\perp}^{max}$ are fixed using the data for the rapidity and the $p_T$ spectra of hadrons.
For example, in the following of this paper, we just use the results obtained by fitting the data of rapidity and $p_T$ spectra of final hadrons
in central Au+Au collisions at $\sqrt{s_{NN}}=$ 200 GeV \cite{BeardMeson04,ptPhenix} with the
aid of the combination model for hadronization\cite{QBXie1988PRD,FLShao2005PRC}.
The results  are   $a=2.40$, $\sigma=2.54$ and $\beta_{\perp}^{max}=0.30$ for $u$ and $d$ newborn quarks
and $a=2.36$, $\sigma=2.73$ and $\beta_{\perp}^{max}=0.34$ for strange quarks.
The numbers of light and strange (anti-)quarks and momentum distribution of net-quarks from the colliding nuclei have been fixed in Ref. \cite{JSong2009MPA}.

\subsection{Charge balance function of the bulk quark anti-quark system}

Having the joint momentum distribution functions, we can calculate the charge balance function in a straight forward way.
In the following, we present the results in rapidity space for different transverse momentum intervals.
In practice, the balance function in rapidity space is often rewritten as a function of the rapidity difference $\delta y=y_{a}-y_{b}$
between two particles in a limited window $y_{w}$, i.e.,
\begin{equation}
B _{ab}(\delta y|y_{w} )=\frac{1}{2} \Bigl\{ \frac{N_{ba}(\delta y,y_{w})-N_{aa}(\delta y,y_{w})}{N_{a} (y_{w} )}+\frac{N_{ab}(\delta y,y_{w})-N_{bb}(\delta y,y_{w})}{N_{b} (y_{w} )} \Bigr \}.
\label{eq:B(deltay)}
\end{equation}

Since quarks of different flavors posses different electric charges, it is not straight forward to extend the definition of the
the electric charge balance function given by Eq.(1) or (\ref{eq:B(deltay)}) to the quark anti-quark system.
There is no direct extension of Eq.(\ref{eq:B(deltay)}) to such cases.
We have many different possibilities at the quark level, e.g.,
\begin{equation}
B_q^{(c1)} (\delta y|y_{w} )= -\frac{1}{2N_f}  \sum_{a,b} \frac{e_ae_b N_{ab}(\delta y,y_{w})}{ e_a^2N_a (y_{w} )} ,
\label{eq:qcorre1}
\end{equation}
\begin{equation}
B_q^{(c2)} (\delta y|y_{w} )= -\frac{1}{2N_f}  \sum_{a,b} \frac{{\rm sgn}(e_ae_b) N_{ab}(\delta y,y_{w})}{N_a (y_{w} )} ,
\label{eq:qcorre2}
\end{equation}
where both $a$ and $b$ run over all the quarks and the anti-quarks, $N_f$ is the number of flavor involved.
We can also defined it as,
\begin{equation}
B_q^{(c3)} (\delta y|y_{w} )=-\frac{1}{2} \Bigl\{
\frac{\sum_{a,b}{\rm sgn}(e_ae_b)N_{ba}(\delta y,y_{w})}{\sum_a N_{a} (y_{w} )}+
\frac{\sum_{a,b}{\rm sgn}(e_ae_b)N_{ab}(\delta y,y_{w})}{\sum_bN_{b} (y_{w} )} \Bigr \},
\label{eq:qcorre3}
\end{equation}
where 
$a=u$, $\bar d$ or $\bar s$ while $b=\bar u$, $d$ or $s$ represent the positively and negatively charged particles respectively.
We may also define the baryon number balance function $B_q(\delta y|y_w)$
for the quark anti-quark system instead, which is given by,
\begin{equation}
B_q^{(b1)} (\delta y|y_{w} )= -\frac{1}{2N_f}  \sum_{a,b} \frac{B_aB_b N_{ab}(\delta y,y_{w})}{ B_a^2N_a (y_{w} )} ,
\label{eq:qcorre4}
\end{equation}
where the summations over $a$ and $b$ run over all different flavors of quarks and those of anti-quarks, and $B_a$ and $B_b$
stand for the baryon numbers.
We can also defined it as,
\begin{equation}
B_q^{(b2)} (\delta y|y_{w} )= -\frac{1}{2} \Bigl\{  \frac{\sum_{a,b}B_aB_b N_{ba}(\delta y,y_{w})}{\sum_{a} B_a^2N_a (y_{w} )}
+\frac{\sum_{a,b}B_aB_b N_{ab}(\delta y,y_{w})}{\sum_{b} B_b^2N_b (y_{w} )} \Bigr\},
\label{eq:qcorre5}
\end{equation}
where $a$ denotes all the quarks of different flavors and $b$ all the anti-quarks of different flavors respectively.
All these definitions satisfy $\int d\delta y B_q(\delta y|y_w)=1$.

We note that, so far as the kind of correlations between the momentum distributions of the quarks and
that of the anti-quarks  described in the working model presented in Sec.A are concerned,
all these definitions do not make much differences.
More precisely,  in the working model presented in Sec. A,  only a correlation between the momentum of the quark and that of the anti-quark
from the same $q\bar q$ pair is introduced as given by Eq.(\ref{eq:covrho}).
There is no correlation between the quarks and anti-quarks from different pairs and there is no difference between different flavors.
In this case, all the definitions given by Eqs.(\ref{eq:qcorre1}-\ref{eq:qcorre5}) are equivalent in the sense that
they are all different suppositions of the correlations given by Eq.(\ref{eq:covrho}) for different flavors and
Eq.(\ref{eq:covrho}) does not distinguish between different flavors.
The only differences come from the net quark contributions where no strange quark exists.

For comparison, we made the calculations using the different definitions Eqs.(\ref{eq:qcorre1}-\ref{eq:qcorre5}) and the results are indeed similar.
In the following part of this section, we show the results obtained by using Eq. (\ref{eq:qcorre3}).

\begin{figure}[!hbtp]
  \center
  \includegraphics[width=0.9\linewidth]{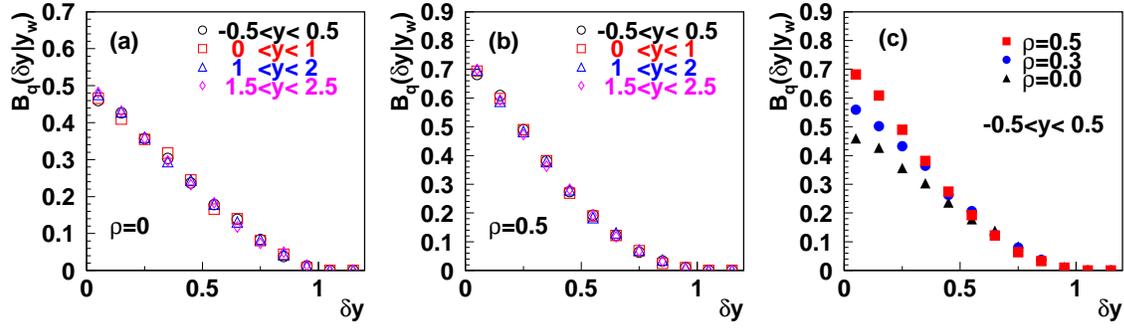}
\caption{The electric charge balance function $B_q (\delta y| y_{w} )$ for the bulk quark system for same window size as a function of $\delta y$ at the variance coefficient $\rho=0$ in panel (a) and $\rho=0.5$  in panel (b), respectively; A comparison of the results at different values of $\rho$ is shown in panel (c).  }
  \label{qbf_rho}
\end{figure}

We first study the case where $\rho=0$. In this case, there is no correlation between the momentum distribution of the quarks and anti-quarks.
The balance is obtained only from the global flavor compensation of the new created quarks and anti-quarks.
This is also the minimum compensation in the produced system.
In Fig. \ref{qbf_rho} (a), we show the results of $y_{w}$=1 for different rapidity positions with transverse momentum $p_\perp$ integrated.
In Fig. \ref{qbf_rho} (b), we show the results for $\rho=0.5$ and a comparison of the results for different values of $\rho$ is given in Fig. \ref{qbf_rho} (c).

From the results, we see that in all cases, also for $\rho=0$, the balance function $B_q (\delta y| y_{w} )$ decrease with increasing $\delta y$ showing a local compensation
of the electric charge in the rapidity space for the bulk quark system.
It is also clear that $B_q (\delta y| y_{w} )$ decreases faster with increasing $\delta y$ for larger value of $\rho$ indicating stronger local charge compensation.
We also see that $B_q (\delta y| y_{w} )$ does not change much for different rapidity window with the same window size showing the longitudinal boost invariance.
This is hold for different values of the variance coefficient $\rho$.

The existence of the approximate boost invariance for the charge balance function for the bulk quark system can easily be understood.
We note that by looking at the different rapidity window in the case that the window size is much smaller than the total rapidity range of the
bulk quark system, we are in fact looking at different fluid cells.
Since we do not differentiate these fluid cells in any significant way, the results should be similar.
This results in similar charge balance function as indicated by the calculated results shown in Fig.  \ref{qbf_rho}.
In other words, the boost invariance of the charge balance function just reflects the homogeneity of the fluid cell at hadronization in different rapidity windows.

\begin{figure}[!htbp]
  \center
  \includegraphics[width=0.9\linewidth]{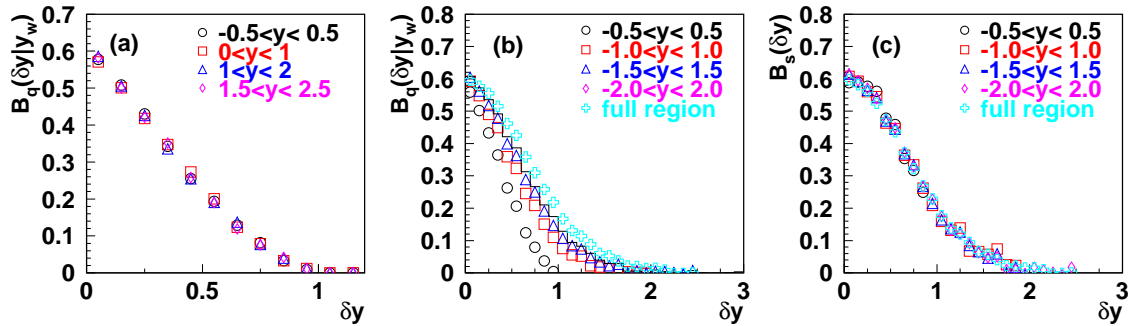}
 \caption{The $p_T$-integrated $B _{q}(\delta y| y_{w} )$ of the constituent quark
system at different rapidity positions with same (panel a) and different (panel b) window
sizes, as well as the $B_{s}(\delta y)$ (panel c).  Correlation coefficient $\rho$ is taken to be 0.3.}
\label{qcorr_pt_integral}
\end{figure}

We continue to study the dependence of the balance function  $B_q (\delta y| y_{w} )$ on the window size and/or transverse momentum.
In Fig. \ref{qcorr_pt_integral} (a), we show   $B_q (\delta y| y_{w} )$ in the different rapidity positions with the same window size $y_{w}=1 $ and
in Fig. \ref{qcorr_pt_integral} (b),  we show   $B_q (\delta y| y_{w} )$ at different window sizes $y_{w}=1, 2, 3, 4 $.
We see that $B_q (\delta y| y_{w} )$ varies with window size and becomes flatter with increasing window size.
This qualitative feature is naturally expected from the definition since the balance function is normalized to unity but
the range of the allowed values of $\delta y$ becomes larger for the larger window size.
This effect can be eliminated by scaling the balance function $B_q (\delta y| y_{w} )$ with the factor $1-\delta y/ |y_w|$
as suggested in Ref \cite{SJeon02}, i.e. we study the scaled balance function,
\begin{equation}
B_{s}(\delta y)=\frac{B _q(\delta y | y_{w})}{1-\delta y/ y_{w} }.
\end{equation}
In Fig.  \ref{qcorr_pt_integral}(c), we show the results obtained for the scaled $B_{s}(\delta y)$ of the bulk quark system.
We see clearly that the scaled balance functions fall on one curve showing that they are independent of the size and position of rapidity window.
For comparison, we also present the balance function in the full rapidity region (open cross) for the case that the net charge of the system is taken to be zero.
We see that the result is also consistent with those for the limited rapidity windows so far as the scaled balance function is studied.
This is very nice feature since it suggests that the scaled balance function for particles in the limited rapidity window can
indeed be regarded as an example for the charge balance function of the system.

We emphasize that these properties of the balance functions of the bulk quark system are results of the momentum distributions of the quarks and anti-quarks in the system .
These distributions including the correlations given by Eq. (\ref{eq:covrho}) are results of the local thermalization and collectivity for the system produced in relativistic heavy ion collisions
in the hydrodynamic theory.
These qualitative features for the charge balance functions for the bulk quark system before hadronization is consistent with those for the final hadrons
as observed by STAR Collaboration at RHIC \cite{StarBF200Scaling}.

\begin{figure}[!htbp]
  \center
  \includegraphics[width=0.9\linewidth]{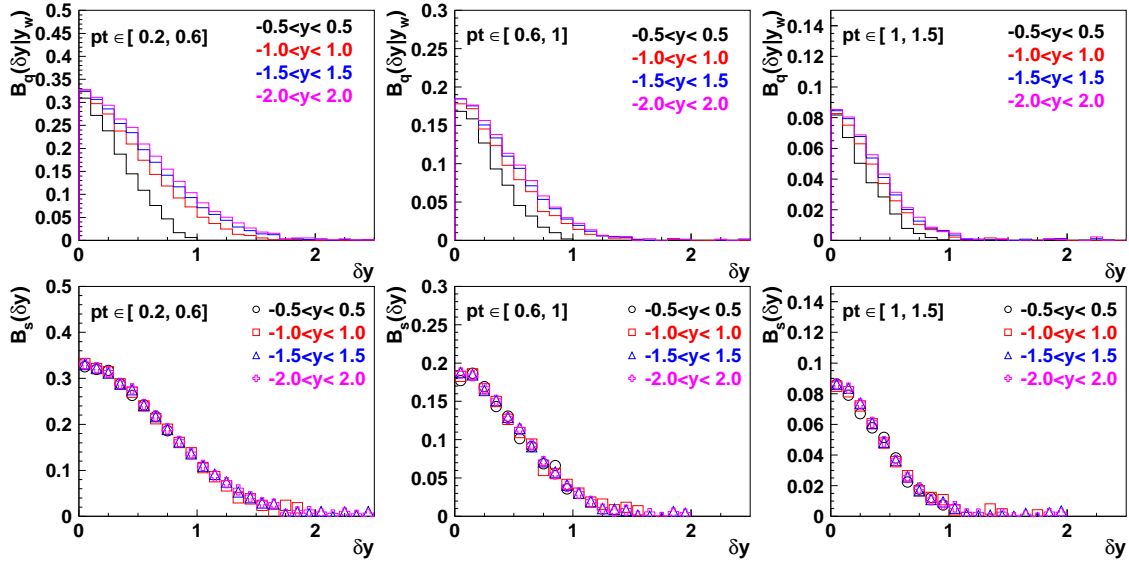}
\caption{The $B_q (\delta y| y_{w} )$ of quark system (top panels)
at different rapidity positions with different window sizes as well as the $B_{s}(\delta y)$ (below panels) in the different $p_T$ (GeV/c) ranges.  Correlation coefficient $\rho$ is taken to be 0.3.}
  \label{qcorr_pt_cut}
\end{figure}

In Fig.\ref{qcorr_pt_cut}, we show $B_q (\delta y| y_{w} )$ and $B_{s}(\delta y)$ in different rapidity windows and in the different $p_{T}$ ranges.
We clearly see that the scaling properties of balance function still hold in the different $p_{T}$ ranges.
We can also see that the width of the scaled balance function decreases  with increasing $p_{T}$.
This is because, in general,  the quarks and anti-quarks with larger $p_T$ come from the fluid cell with larger transverse flow,
which results in a smaller longitudinal rapidity interval and hence smaller width for balance function.
Such a feature was expected earlier at the hadron level\cite{Bialas04} and observed in
central Au+Au collisions at $\sqrt{s_{NN}}= 200$ GeV \cite{StarBF200Scaling}.

\section{charge balance functions of the hadron system}

With the momentum distribution functions of the bulk quark system discussed in last section,
we study the charge balance functions of hadrons produced in the hadronization of this system.
We compare the results obtained for the directly produced hadrons and those of the final state hadrons with those for the quarks and anti-quarks
to study the influence of the hadronization and resonance decay on the balance functions.

We describe the hadronization of the bulk quark system with the (re-)combination or coalescence mechanism.
Such a hadronization mechanism is tested by various data and is implemented in different forms such as
the quark recombination model \cite{Fries2003PRL,RCHwa04},  the parton coalescence model \cite{Greco2003PRL,Molnar2003PRL},
and the quark combination mode (SDQCM) \cite{QBXie1988PRD,FLShao2005PRC}.
All these models are tested against the various features of the hadrons produced in heavy ion collisions at high energies.
Here, in this paper, we use SDQCM \cite{QBXie1988PRD,FLShao2005PRC} for our calculations
since this model  takes the exclusive description and is implemented by a Monte-Carlo program so that can be apply to calculate
the balance functions for the directly produced hadrons as well as the final hadrons after the resonance decays in a very convenient way.
Also, this model guarantees that mesons and baryons exhaust all the quarks and anti-quarks in the deconfined color-neutral system at hadronization.

\subsection{Charge balance functions in rapidity space}

We insert the momentum distributions including the correlations given by Eq. (\ref{eq:covrho}) to determine the momenta of the quarks and anti-quarks before hadronization.
We then apply the quark combination rules as implemented in the Monte-Carlo program of SDQCM\cite{FLShao2005PRC} to calculate the momentum distribution of the directly produced hadrons.
Those resonances will decay accordingly and the momentum distributions are simulated also in the program by using the material from the particle data group\cite{pdg08p355}.

\begin{figure}[!htbp]
  \center
  \includegraphics[width=0.9\linewidth]{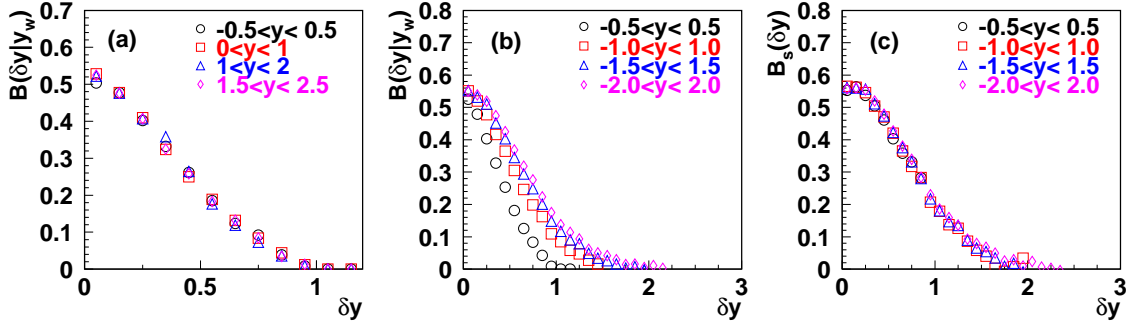}\\
  \caption{The $p_T$-integrated $B (\delta y| y_{w} )$ of initial
  hadron system at different rapidity positions with same (panel a) and different (panel b) window
sizes, as well as the $B_{s}(\delta y)$ (panel c). Correlation coefficient $\rho$ is taken to be 0.3.}
\label{ini_hadron_window}
\end{figure}

\begin{figure}[!htbp]
  \center
  \includegraphics[width=0.9\linewidth]{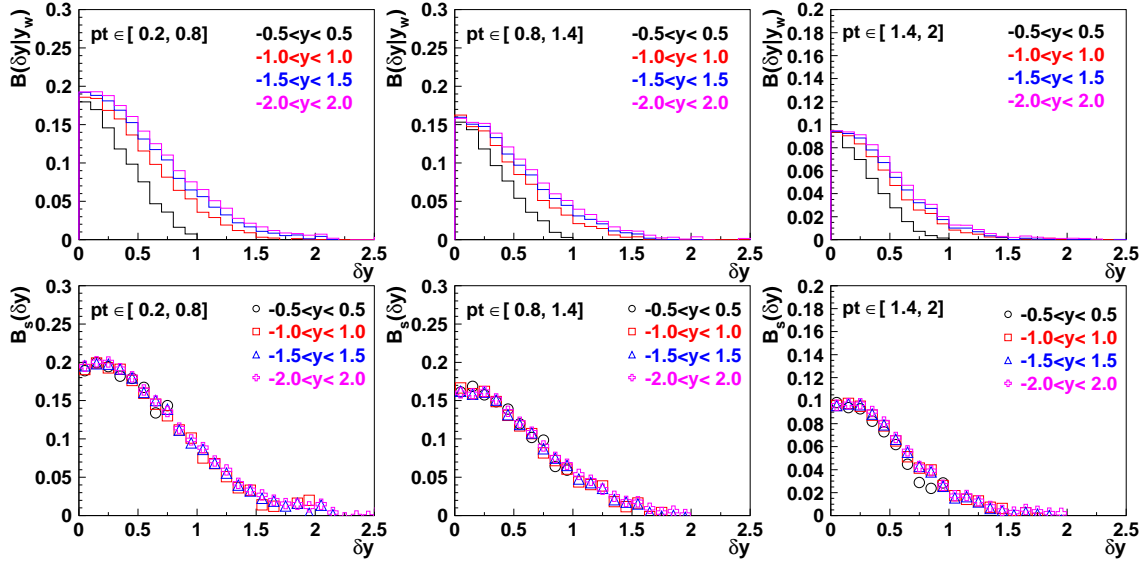}
\caption{The $B (\delta y| y_{w} )$ of initial hadron system (top
panels) at different rapidity positions with different window sizes as well as the $B_{s}(\delta y)$ (below panels) in the different $p_T$ (GeV/c) ranges. Correlation coefficient $\rho$ is taken to be 0.3.}
  \label{ini_hcorr_pt_cut}
\end{figure}

In Fig. \ref{ini_hadron_window}, we show the results for the $p_T$-integrated balance functions for the directly produced hadrons.
Here, in  Fig. \ref{ini_hadron_window}(a), we see the results in different rapidity windows with the same width $y_{w}=1$,
while in Fig. \ref{ini_hadron_window}(b) and (c), we see the results at different window sizes as well as the scaled function $B_{s}(\delta y)$.
In Fig.\ref{ini_hcorr_pt_cut}, we show the corresponding results in different $p_T$ ranges.

From these results, we see that both the longitudinal boost invariance and rapidity scaling for the balance functions are hold for the
hadrons directly produced in the quark combination mechanism, either for the $p_T$-integrated quantities or those for different $p_T$ ranges.
This is in fact not surprising because the formation of hadrons in this hadronization mechanism is realized by the combination of two
or three nearest quarks/antiquarks in momentum space.
This means that the combination happens locally and does not destroy the locality nature of charge balance of the system.

\begin{figure}[!htbp]
  \center
  \includegraphics[width=0.9\linewidth]{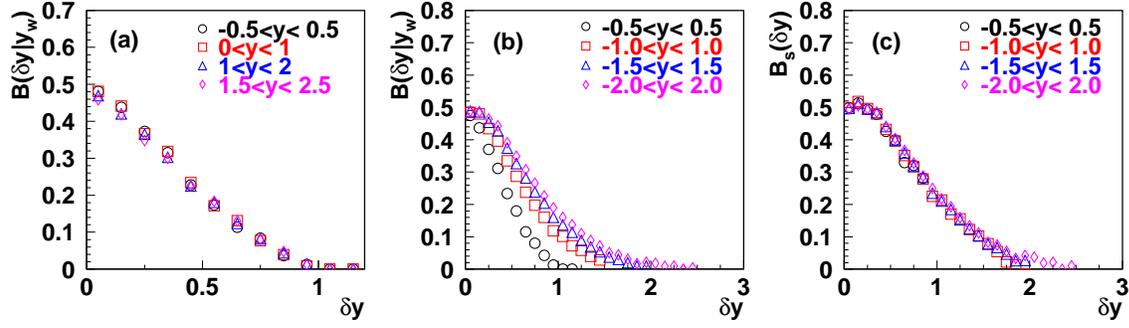}\\
  \caption{The $p_T$-integrated $B (\delta y| y_{w} )$ of final
  hadron system at different rapidity positions with same (panel a) and different (panel b) window
sizes, as well as the $B_{s}(\delta y)$ (panel c). Correlation coefficient $\rho$ is taken to be 0.3.}
\label{final_hadron_window}
\end{figure}

\begin{figure}[!htbp]
  \center
  \includegraphics[width=0.9\linewidth]{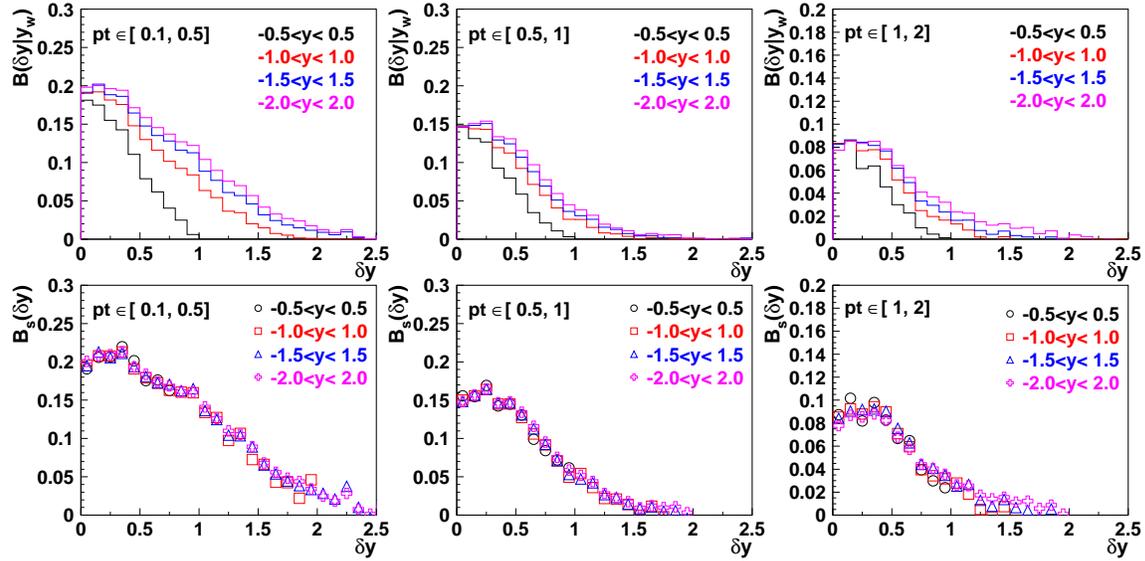}
\caption{The $B (\delta y| y_{w} )$ of final hadron system (top
panels) at different rapidity positions with different window sizes as well as the $B_{s}(\delta y)$ (below panels) in the different $p_T$ (GeV/c) ranges. Correlation coefficient $\rho$ is taken to be 0.3.}
  \label{final_hcorr_pt_cut}
\end{figure}

We further study the resonance decay contributions by calculating the balance functions for the final hadrons
where decays of the resonances are taken into account.
We show the corresponding results in Figs. \ref{final_hadron_window} and  \ref{final_hcorr_pt_cut}.

From Figs. \ref{final_hadron_window} and  \ref{final_hcorr_pt_cut}, we see that both the boost invariance in rapidity space and the scaling property
still preserved after the contributions from the resonance decays are taken into account.
Together with those results given in Figs. \ref{ini_hadron_window} and \ref{ini_hcorr_pt_cut}, these results show clearly,
although there are definitely influences from hadronization and resonance decay on the form of the charge balance functions,
these effects do not significantly influence the boost invariance and the scaling in rapidity space.

The influences from hadronization and resonance decay to the balance function can be studied more quantitatively by calculating the averaged width of the balance function, which is defined as,
\begin{equation}
 \langle \delta y \rangle = \frac{\int\nolimits_{0}^{y_{w}} B (\delta y| y_{w} )\ \delta y \ d\delta y}
 {\int\nolimits_{0}^{y_{w}} B (\delta y| y_{w} ) d\delta y}.
\end{equation}
We note that the averaged width $\langle \delta y \rangle$ is in general a charactering quantity describing the radius of charge balance of the system.
For the final hadron system in heavy ion collisions, it can be sensitive to different effects such as
delayed hadronization or hadron freeze-out \cite{Bass_clockHd,FuQHbf11}, possibly highly localized charge balance at freeze-out \cite{Schlichting11_bf},
transverse flow \cite{Bialas04,bozek05}, multiplicity effect \cite{NA22BFkp,DuJX07} and hadronic weak decay.
Here, by comparing the results for the balance functions of the quark anti-quark system with those for the initial hadrons and those for the final hadrons,
we can study the magnitudes of the influences of hadronization and those from resonance decay.

\begin{figure}
  \includegraphics[width=\linewidth]{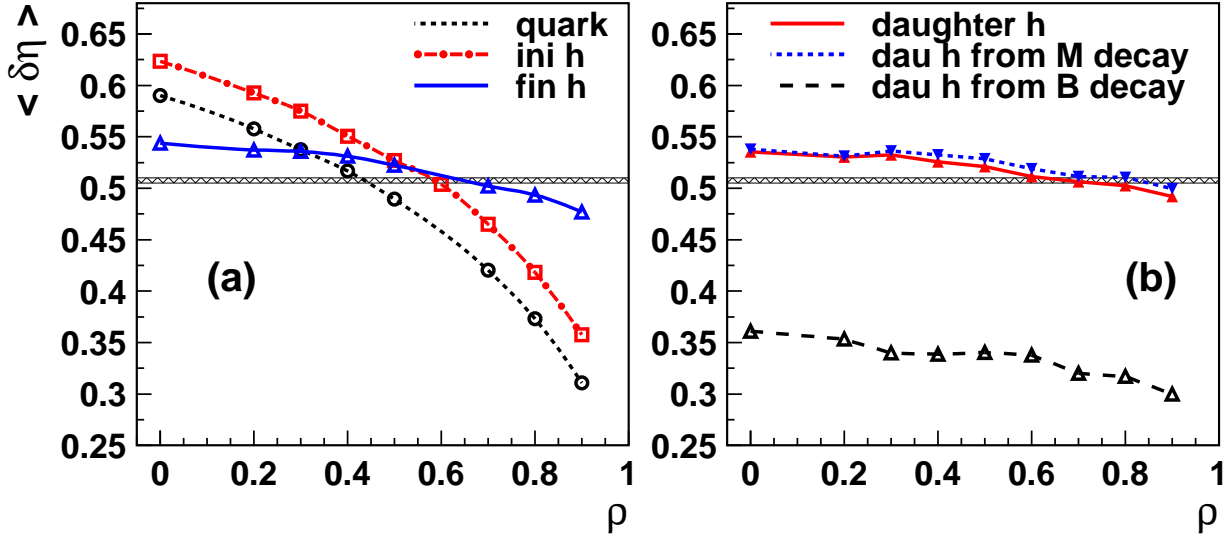}\\
  \caption{In (a), we see the averaged widths $\langle \delta \eta \rangle$ of the balance functions for
  the quark and anti-quark, the directly produced and the final hadron systems as functions of $\rho$.
  In (b), we see the averaged widths $\langle \delta \eta \rangle$ of the balance functions for
  the daughter hadrons from all the resonance decays, those for the daughter hadrons from the meson decays and those from the baryon decays, respectively.
  The band area represents the experimental of $\langle \delta \eta \rangle$ for charged particles in central Au+Au $\sqrt{s_{NN}}=$ 200 GeV \cite{StarBF200Scaling}. }
  \label{bfetaw}
\end{figure}

In Fig. \ref{bfetaw}(a), we show the results for the averaged width $\langle\delta \eta\rangle$ of the balance function for the directly produced hadrons as the function of $\rho$,
compared with that for the quarks and anti-quarks and that for the final hadrons.
Here, we show the results for pseudo-rapidity $\eta$ in order to compare with the experimental data now available \cite{StarBF200AApp}.
We choose also the same pseudo-rapidity and $p_T$ regions as the experiments \cite{StarBF200AApp}, i.e. $|\eta|<1$, $0.1\le |\delta \eta| \le 2.0$ and $0.2<p_{T}<2.0$ GeV/c.
In Fig. \ref{bfetaw}(a), the data of $\langle \delta \eta \rangle$ in central Au+Au collisions at $\sqrt{s_{NN}} =$ 200 GeV \cite{StarBF200AApp} is shown as a band area.
We see clearly that, in all the three cases,  the averaged width $\langle \delta \eta \rangle$ decreases with the increasing $\rho$.
We see also that,  the $\langle \delta \eta \rangle$ for the directly produced hadron system decreases with the increasing $\rho$ in exactly the same way as that for the bulk quark system does.
The difference between $\langle \delta \eta\rangle$ for the directly produced hadron system and that for the bulk quark system is almost a constant 0.04 for all the different values of $\rho$.
This is because, as mentioned above, the combination of quark(s) and/or anti-quark(s) in neighbor does not change the electric charge balance in any essential way.
However, the formation of electrically neutral hadrons may delay the charge balance in momentum space.
We take a quark and an anti-quark from a given $q\bar q$-pair as an example.
As given by Eq.(\ref{eq:covrho}), their momentum distributions posses a correlation measured by the covariance coefficient $\rho$.
If both of them enter into the respectively charged hadrons in the combination process,  the correlation will pass to the hadronic level.
However,  if one of them enters into an electrically neutral hadron, the correlation will be lost in the charged hadrons.
This will decrease the local charge balance at the hadron level.

From Fig. \ref{bfetaw}(a), we also see that the resonance decay contributions change $\langle \delta \eta \rangle$ significantly.
It is also interesting to see that these contributions strengthen the local charge balance for relatively small values of $\rho$ but
weaken the balance for larger values of $\rho$.
This indicates that the decay contributions dilute the balance functions quite significantly.
To see where these different behaviors come from, we calculate the averaged width $\langle \delta y\rangle$ for those hadrons from
resonance decay separately.
We note that the influence of resonance decay to the charge balance function is in general different for hyperon decay from that for vector meson decay.
The decay of the hyperons such as $\Lambda \to p \pi$ and $ \Xi^{0} \to \Lambda \pi$ produces a pair of charged daughter particles
with quite narrow rapidity interval, e.g. about one third for  $\Lambda \to p \pi$ and $ \Xi^{0} \to \Lambda \pi$,
due to the small amount of energy released in the decay process.
This leads also to a smaller $\langle \delta \eta \rangle$ for the charge balance function.
However, in vector meson decay such as $\rho^0\to\pi^+\pi^-$ and $K^{*0}\to K^+\pi^-$,
the energy released is much larger leading to a much larger rapidity difference between the daughter particles,
e.g. up to 1.7 for $\rho^0\to\pi^+\pi^-$ and 1.3 for $K^{*0}\to K^+\pi^-$ in the rest frame of parent particle.
To study this effect in a more quantitative manner, we calculate the averaged width  $\langle \delta \eta \rangle$  of the balance function
only for the daughter particles from baryon or meson decay respectively.
The results are shown in Fig. \ref{bfetaw} (b).
Here, we see clearly that, the averaged width  $\langle \delta \eta \rangle$ for the daughter particles from baryon decay
is indeed much smaller than those from meson decay.
We see also that the charge balance for the daughter particles from baryon decays is dominated by the decay effect
which leads to an averaged width of about one third.
However, for those from meson decays, the charge balance is dominated by the effect from the mother particles.

\subsection{Charge balance in the azimuthal direction}

The charge balance in the azimuthal direction for hadrons in high energy heavy ion collisions can be sensitive to jet production.
Experimental studies have already been carried out by STAR Collaboration for hadrons of  different  $p_T$ regions \cite{StarBF200AApp}.
It is thus also interesting to see how the charge balance function behaves for the bulk quark matter system and the resulting hadrons.

The balance function of hadrons in the azimuthal direction is defined similarly to that in rapidity,
\begin{eqnarray}
B_{ba,azi}(\delta\phi,\phi) = \frac{1}{2} \left\{
\frac{ N_{ba}(\delta \phi,\phi)  - N_{aa}(\delta\phi,\phi)}{  N_a (\phi) } 
+
{ N_{ab}(\delta\phi,\phi)  -  N_{bb}(\delta\phi,\phi)  \over
 N_b (\phi)} \right\},
\label{defphi}
\end{eqnarray}
where, e.g.  the  quantity $N_{ba}(\delta \phi,\phi)$ is  defined as the number of pairs  of particles with the
particle $a$ flying at an angle $\phi$  (measured with respect to the reaction plane) and
the particle ${b}$  at an angle between $\phi$ and $\phi+\delta\phi$.
In the following we study the azimuthally averaged balance function
\begin{equation}
B_{ab,azi}(\delta\phi)=\int_0^{2\pi} d\phi B_{ab}(\delta\phi,\phi).
\end{equation}

\begin{figure}
  \includegraphics[width=\linewidth]{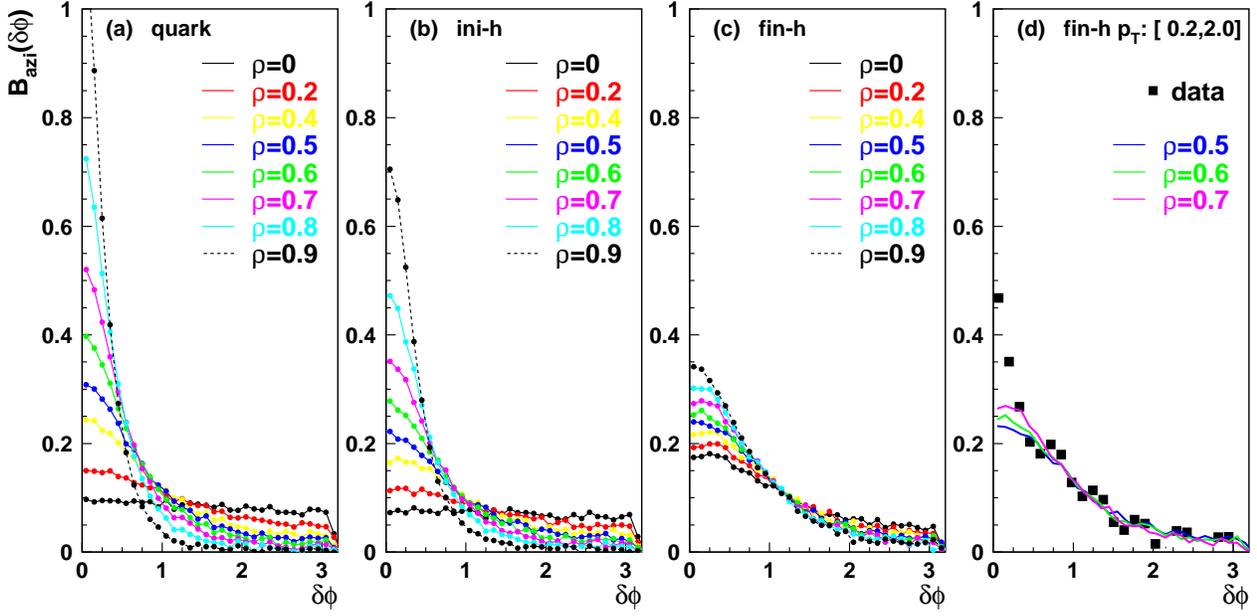}\\
  \caption{Balance function $B_{azi}(\delta \phi)$ of the quark anti-quark system (a),
  the directly produced hadron system (b) and the final hadron system (c) as functions of $\delta \phi$
  in the pseudo-rapidity region $-1<\eta<1$ and $0.2<p_T<2.0$GeV/c.
  A comparison of the results for the final hadron system at $\rho= 0.5, 0.6 $ and 0.7 and
 the experimental data for all charged particles with  $0.2 < p_T < 2.0 $GeV/c in
 central Au+Au collisions at $\sqrt{s_{NN}}=$ 200 GeV in Ref.\cite{StarBF200AApp}
 is given in (d).  }
 \label{delt_phi}
\end{figure}

Having the Monte-Carlo program at hand, the extension of the calculations mentioned above to azimuthal direction is straight forward.
We show the results obtained for the quark, the directly produced hadron and the final hadron system at different $\rho$ values in Fig. \ref{delt_phi}.
Comparison with the available data\cite{StarBF200AApp} is also given in the figure.

From the results for the bulk quark system Fig. \ref{delt_phi}(a), we see clearly that the dependence of the quark charge balance function on the
variance parameter $\rho$ is quite obvious.
For $\rho$ close to unity, the momentum of the quark and that of the anti-quark produced in pair are closely correlated, and we see a sharp peak at $\delta\phi=0$.
For $\rho=0$, there is no correlation and the balance function is almost a flat function showing only the influence from the global charge compensation.

Comparing the results in Fig. \ref{delt_phi}(b) with those in Fig. \ref{delt_phi}(a),
we see that the charge balance functions in the azimuthal direction for the directly produced hadron system
are slightly broader than the corresponding results for the quark system, showing a slightly loose correlation.
This is similar to the case in rapidity space studied in last subsection.
However, the influences from the resonance decays are quite significant in the azimuthal direction.
We see quite significant differences between the results for the final hadrons and the corresponding results for the hadron system before resonance decay.
We see in particular that the very much pronounced peak at $\delta\phi=0$ is smoothed by the decay influences.
This is also obvious since such strong correlation can be destroyed by the resonance decay because of the
transverse momentum conservation in the decay processes.
From Fig. \ref{delt_phi}(c), we see that  the data\cite{StarBF200AApp} is well be described except for the peak at $\delta\phi=0$.
This peak could be an indication of jet contribution which is not included in our calculations.

\section{summary}
In summary, we have calculated the charge balance functions of the bulk quark system before hadronization,
those for the directly produced and the final hadron system in relativistic heavy ion collisions.
The momentum distributions for the quarks and the anti-quarks in the bulk quark system are taken as
determined in the hydrodynamic picture with local thermalization and collectivity.
A correlation between the momentum distribution of the quark and that of the anti-quark is introduced if they are from the same new produced $q\bar q$ pair and
the correlation strength is described by the variance coefficient $\rho$.
Our results show that the charge balance functions for the bulk quark system have the longitudinal boost invariance and the scaling behavior in rapidity space.
Such properties are preserved by the subsequent hadronization via combination mechanism and the resonance decay,
although both hadronization and resonance decay can influence the width of the balance function.
With the same inputs, we also studied the balance function in the azimuthal direction.
\section*{ACKNOWLEDGMENTS}
The authors thank Q. B. Xie, Q. Wang and G. Li for helpful discussions.  The work is supported in part by the National Natural Science Foundation of China under grant 11175104, 10947007, 10975092, and by  the Natural Science Foundation of Shandong Province, China under grant ZR2011AM006.

\end{document}